\documentclass[12pt,preprint2]{aastex}
\usepackage{emulateapj5,apjfonts,epsfig,onecolfloat5}

\newcommand{\alt}{\lesssim}
\newcommand{\agt}{\gtrsim}

\slugcomment{\today}

\shorttitle{Counterparts to gravitational wave events}
\shortauthors{Sylvestre, J.}

\begin{document}

\twocolumn[
\title{Prospects for the detection of electromagnetic counterparts to gravitational wave events}
\author{Julien Sylvestre}
\affil{LIGO Laboratory, California Institute of Technology}
\affil{MS 18-34, Pasadena, CA 91125, USA.}
\email{jsylvest@ligo.caltech.edu}

\begin{abstract}
Various models for electromagnetic emissions correlated with the gravitational wave signals expected to be detectable by the current and planned gravitational wave detectors are studied.
The position error on the location of a gravitational wave source is estimated, and is used to show that it could be possible to observe the electromagnetic counterparts to neutron star-neutron star or neutron star-black hole binary coalescences detected with the Advanced LIGO and the Virgo detectors.
\end{abstract}

]

\section{Introduction}
A number of large laser interferometric detectors of gravitational waves (GW), developed by the LIGO project in North America, the GEO and the Virgo projects in Europe, and the TAMA300 project in Asia, are rapidly approaching their sensitivity goals.
Months long data taking runs with three or more detectors are planned or have been completed in 2003.
In addition, a significant upgrade of the LIGO project interferometers which is planned for the end of the decade, the Advanced LIGO system, will lead to a ten-fold improvement in sensitivity in these detectors.
Operating these GW detectors in a tightly coupled network has the advantage of reducing the likelihood of false detections, or equivalently of achieving a better distance reach for a given false alarm rate.
The network can also be used to infer the position of the source.
I argue in this paper that the precision of the position estimation and the electromagnetic (EM) fluxes expected from the most easily detectable sources of GW should be marginally sufficient to allow the observation of EM counterparts to GW events.

Current estimates suggest that the most likely GW signal to be observed is the ``chirp'' from the in-spiral of two compact objects (neutron stars [NS], or black holes [BH]\footnote{Wherever it is relevant, the mass of NS is assumed to be $1.4 M_\odot$, and the mass of BH, $10 M_\odot$.}) in a close binary.
A single (Advanced) LIGO detector should achieve a detection rate for NS-NS and NS-BH compact binaries of $2\times 10^{-3} - 0.7$ yr$^{-1}$ ($10 - 10^3$ yr$^{-1}$) [\citet{B2001}].
These ranges correspond to a number of binary formation models; the ``Standard model'' of \citep{B2001} gives a NS-NS (NS-BH) coalescence rate of 50 Myr$^{-1}$ (10 Myr$^{-1}$) in the Milky Way.
For NS-NS binaries, the signal-to-noise ratio (SNR) in a single LIGO detector (Advanced LIGO detector), assuming optimal signal processing, will be 10 for an optimally oriented source at 25 Mpc (425 Mpc), and will scale inversely with the source distance \citep{Finn1993}.
Averaging over source position and orientation reduces the distance for fixed SNR by a factor of $\sim 5/2$.
However, a network of three similar, independent detectors would achieve the same SNR at a distance larger by a factor of $\sim \sqrt{3}$.

Only limited attention has been given to the important problem of understanding the EM emissions of compact binary mergers.
I review in section \ref{EM} a few mechanisms which may be important for such emissions.
I then present in section \ref{GW} an overview of the techniques which will be used to analyze the GW data in order to discover and locate coalescing compact binaries.
A comparison of the localization performances of these techniques and of the observational capabilities of EM detectors is presented in section \ref{counterparts}, and is used to estimate the observation rate of counterparts to compact binary mergers, for various networks of interferometers.

\section{Electromagnetic signals} \label{EM}
I review below three different mechanisms that may be important in generating EM emission when a compact binary coalesces: magnetospheric interactions, the radioactive decay of ejected material, and relativistic blast waves.
These models are rather crude in their predictions, and it is not clear which, if any, may accurately describe actual EM counterparts.
They should be used below as order-of-magnitude estimates of the EM signal.
Alternatively, the observational prospects quoted below can be interpreted as prospects for setting upper limits on these models from the EM observation of GW events.

When a NS orbits a strong magnetic field companion, an electric field is induced in the orbiting star, leading to particle acceleration in the form of a stellar wind, and coherent EM radiation as in normal pulsars.
\citep{Hansen2001} predict that this results in a radio burst {\it precursor} occurring seconds before the GW burst from the merger and during the in-spiral, with flux at 400 MHz of $F \sim 2.1 \;{\rm mJy} ( r / 100 \;{\rm Mpc})^{-2} ( B/10^{15} \;{\rm G})^{2/3},$
for $r$ the distance to the source, and $B$ its magnetic field intensity.
In addition, a large amount of energy is extracted from the orbital motion and released in the magnetosphere as Alfv\'en waves and a pair plasma.
This energy could drive a relativistic wind of pairs and photons, which would become optically thin after some expansion (still before the binary merger), and yield a X-ray thermal emission lasting a few seconds with flux 
$F \sim 3\times 10^{-9} \;{\rm erg}\;{\rm cm}^{-2}\; {\rm s}^{-1} ( r/100 \;{\rm Mpc})^{-2} ( B/10^{15} \;{\rm G})^2,$
and temperature increasing from 10 keV to 100 keV during the burst, according to the same authors.
The fact that the magnetospheric interaction model leads to bursts that occur before the end of the GW signal is problematic for the X-ray burst, but, as pointed out by \citep{Palmer}, interstellar dispersion could delay the radio burst enough to allow its observation.
Alternativelly, if X-ray or radio instruments monitoring most of the sky are available, archival searches for coincidences with GW signals could be possible.

In the second scenario, it is the radioactive decay of the neutron rich nuclei of the decompressed NS matter ejected during the merger that produces the energy required to power an EM signal.
Numerical simulations \citep{Rasio1999} suggest that a mass shedding instability ejects $\sim 10\%$ of the NS mass after one or two orbits following first contact.
\citep{Paczynski1998} consider a simple model where some of this matter expands in a spherical envelope of mass $M$, which is heated by the radioactive decay of the ejecta.
If a fraction $f$ of the envelope mass decays and is converted to heat, they predict a peak luminosity
$L \sim 2\times 10^{44} \;{\rm ergs}\;{\rm s}^{-1} (f/10^{-3}) (M/0.01 \; \rm{M}_\odot )^{-1/8}$
and an effective temperature at peak luminosity around $3\times 10^4 \;{\rm K}(M/0.01 \;\rm{M}_\odot)^{1/8}(f/10^{-3})^{1/4}$, so that most of the EM radiation will be emitted as soft UV.
The optical remnant should decay on a timescale of $\sim 1$ day.

Finally, a third possibility is that the merger of a compact binary generates a relativistic blast wave, as it has been argued to explain long wavelength counterparts to gamma-ray bursts [e.g., \citet{vpal2000}].
The black hole accretion disk model \citep{MR1997} predicts that NS-NS or BH-NS coalescences result in an intermediate state where an accretion disk forms around a remnant BH, and that this accretion disk is responsible for the jet that powers the gamma-ray burst.
The disk is accreted on viscous timescales, so that this model is likely to be useful only for short gamma-ray bursts [duration $\alt$ 1 s, \citet{NPK2001}].
Gamma-ray bursts, when seen on-axis if they are non-isotropic, are visible at cosmological distances, i.e. much farther out than where the GW signal from binary coalescences will be visible; for such bright EM sources, simple time coincidences with GW detections should be sufficient to identify the EM counterpart \citep{KM2002}.
However, depending on the currently uncertain amount of beaming in short gamma-ray bursts, a possibly large fraction of GW events might only have ``orphan'' afterglows with no detectable gamma-ray emission \citep{KM2002}: the afterglow is produced by the decelerating outflow, so that the relativistic beaming of its radiation is decreasing with time, thus making the emission more and more isotropic, and thus more likely to be observed.
It might also be that some compact binary mergers generate a fireball that powers an afterglow, but fails to generate gamma-ray bursts \citep{huang2002}, so the search for counterparts to GW bursts should not be limited to gamma-ray bursts.

The detection of afterglows for short duration gamma-ray bursts appears to be much more difficult than for long duration gamma-ray burst.
This could be due to larger error boxes for short bursts, or to intrinsically weaker afterglows.
\citep{PKN2001} argue for the latter hypothesis, scaling a model for long duration bursts to short duration ones by taking the energy of the short burst to be $5\times 10^{51}/4\pi$ ergs sr$^{-1}$, and the density of the surrounding medium to be low, $10^{-3}$ cm$^{-3}$, since binary mergers are likely to occur outside their parent galaxy.
For a source at $z=1$, the R band magnitude, 0.1 day after the collapse, is then $R \sim 24$, and the 2-10 keV flux is $F_{2-10k} \sim 3\times 10^{-13} \;{\rm erg}\;{\rm cm}^{-2}\;{\rm s}^{-1}$.
The radio flux peaks $\sim 3$ days after the collapse, and is $\alt 3 \times 10^{-3}$ mJy\footnote{In extrapolating these numbers to smaller distances below, I assume a power law dependence of the optical and X-ray flux on the radiation frequency, $F\propto \nu^{-p/2}$, for an electron energy index $p\sim 2$, and a synchrotron slope $F\propto \nu^{1/3}$ for radio waves.}.

Another source of GW with a well understood EM counterpart is the collapse of the core of massive stars in a supernova explosion.
According to recent estimates \citep{Fryer2001}, a good channel for GW emission in stellar collapses is the excitation of a bar-mode instability in newborn, rapidly rotating neutron stars.
From the data of (\citet{Fryer2001}, Fig. 7), the collapse of a 15$M_\odot$ star would produce a detection with SNR in a single Advanced LIGO detector of $\sim 8 (10 {\rm Mpc}/r)(N_c/100)$, where $N_c$ is the number of cycles over which the bar is assumed to remain coherent, and where optimal signal processing and orientation are assumed. 
The rate of core-collapse supernovae in the Galaxy is well-known, and lies between 0.007-0.02 per year.
The fraction of all core-collapse supernovae that result in a neutron star with enough spin to develop the bar-mode instability is, however, largely unknown.
Assuming this fraction to be one, averaging the response of the detector over the whole sky, and using the extragalactic rate extrapolation method of \citep{Kalogera2001}, this gives an upper limit on the detection rate of $0.3 (N_c/100)^{3}$ per year (SNR $\agt 5)$.
The mean maximum absolute B magnitude of Type II supernovae is -16.9, with standard deviation 1.4 \citep{Miller1990}, so that the EM emission is detectable ($B \alt 20$, assuming no reddening) for any GW detection out to 10 Mpc.
It should be noted that \citep{Dahlen} predict $\sim 0.36$ core collapse supernovae per square degree with $R > 22$, at any given time.
For the farthest supernovae detectable using GW, supernovae unrelated with the GW signal might be observed optically within the source position error box.

\section{Localization of the gravitational wave source} \label{GW}
The interest in the problem of efficiently detecting GW bursts in real data with a significant non-Gaussian noise component is currently driven by the availability of preliminary data from the LIGO and the GEO projects [e.g., \citet{thesis}].
Models of the data analysis algorithms that use Gaussian noise are very useful in getting good estimates of the performances of these algorithms, and should become more accurate descriptions of the real analysis as the quality of the data improves.
With this caveat, it is currently well understood how the analysis of the data for GW signals of a precisely known form should be performed: given a set of parameters describing the signal, a template waveform is formed, and is correlated with the data.
The full parameter space of the signal is explored with a finite number of points chosen so that the mismatch between a true signal and its closest approximation leads to a reduction in SNR that is smaller than a few percents.
This procedure, {\it matched filtering}, can be applied independently or coherently to geographically distributed detectors.
In the former case, an estimate of the signal parameters, including its arrival time, is obtained at every detector, and the source position is estimated by triangulation, using the relative phase of the signal in all detectors.
In the latter case, the response to a template waveform is calculated coherently using the data from all detectors (i.e., by adding the log-likelihood from all interferometers for a single set of source parameters), and the position estimate is given by the point in parameter space with the largest response.
The coherent approach is more efficient in terms of detection efficacy and error boxe sizes \citep{Pai2000}, but it might be computationally prohibitive to implement, at least in its most naive form \citep{Pai2001}.

In the results given below, I will consider the network of three interferometers consisting of the Virgo ($V$) instrument near Pisa (Italy), the LIGO instrument near Livingston ($L$), Louisiana (USA), and the 4 km LIGO instrument near Hanford ($H$), Washington (USA).
In the case of triangulation, the form of the position error, in terms of the solid angle $\Delta\Omega$ containing the true source position 95\% of the time, for experiments repeated on independent data but for the same true source position, is [Kip Thorne, cited in \citet{tinto}]:
\begin{equation}
\Delta\Omega \propto \frac{c^2 \Delta t_{LV} \Delta t_{HV}}{A_{HLV} |\cos\theta|}, \label{eq:dOmega}
\end{equation}
where $\Delta t_{ij}$ is the standard deviation on the time delay between interferometers $i$ and $j$ (the $HL$ baseline can also be used if it reduces the position error), $A_{HLV}$ is the area of the triangle formed by the three interferometers, and $\theta$ is the angle between the normal to that triangle and the direction of propagation of the GW.
Assuming Gaussian errors, $(\Delta t_{ij})^2 = (\Delta t_i)^2 + (\Delta t_j)^2$, for $\Delta t_i$ the standard error on the signal's arrival time estimation in detector $i$.
The SNR and $\Delta t_i$ are calculated for each interferometer for a ``Newtonian chirp'' for NS-NS and NS-BH coalescences as in \citep{Finn1993}, but with up-to-date noise spectra\footnote{Initial LIGO: www.ligo.caltech.edu/$\sim$kent/ASIS\_NM/noise\_models.html; Virgo: www.virgo.infn.it/senscurve; Advanced LIGO: LIGO Document LIGO-M990288-A-M.}.
I have calibrated Eq. (\ref{eq:dOmega}) using numerical simulations, so that the proportionality constant multiplying the right-hand side of the equation is $\sim 5.0$.

The position error from a detection of a NS-NS coalescence using the coherent approach is taken from \citep{Pai2000}, and is given by
\begin{equation}
\Delta\Omega = \frac{2\times 3.7\times 10^{-4} \;{\rm sr}}{|\cos \theta|} \left(\frac{12}{\rho_N}\right)^2, \label{eq:coherent}
\end{equation}
where $\rho_N$ is the {\it network} SNR, which is the sum in quadrature of the SNRs in all detectors.
I include the extra factor of 2 to convert the quoted $1\sigma$ result to a $2\sigma$ error box (95\% probability coverage).
With the initial network, the coherent method yields an error box the same size as triangulation for a source at $\theta=0$, but a factor of $\sim 3.6$ farther (in good agreement with $\Delta\Omega \propto {\rm SNR}^{-2}$). 
Networks with detectors having characteristics similar to those of the Advanced LIGO detectors have not been studied by \citep{Pai2000}, so I will interpolate their results by using Eq.\ref{eq:coherent} with the network SNR computed for the Advanced LIGO noise characteristics, when detectors at all sites (including Virgo) are assumed to have noise performances close to the Advanced LIGO design.
I will also consider the intermediate network consisting of two Advanced LIGO detectors and of the Initial Virgo detector.
In that case, the error box will be an elongated ellipse, instead of being mostly circular.
To account for this asymmetry, I use $\rho^2_N = \rho_V\sqrt{\rho^2_{H} + \rho^2_L}$ in Eq. (\ref{eq:coherent}), where the SNRs for the LIGO instruments are computed with Advanced LIGO noise power spectra, and the SNR for Virgo uses its Initial noise spectrum.
The advantages of using the coherent analysis are not as large for this network as for networks with similar instruments at all sites.

\section{Observation prospects for counterparts} \label{counterparts}
\begin{figure}
\plotone{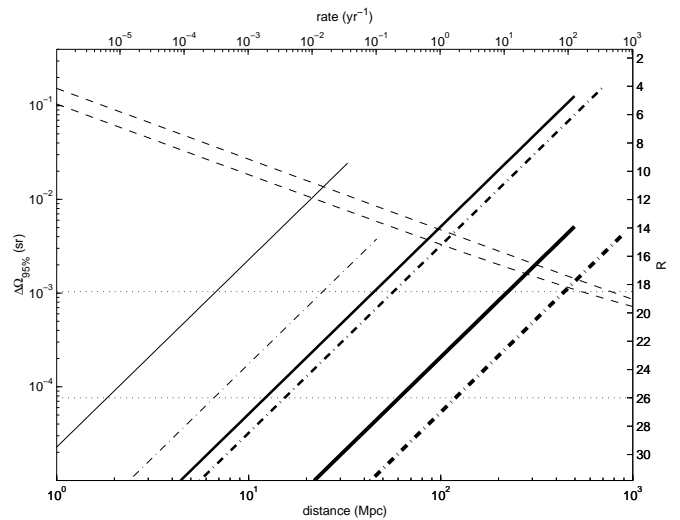}
\caption{The detectability of optical counterparts to the coalescence of a NS-NS binary, as a function of the luminosity distance of the source. The continuous lines show the expected position error (left-hand side axis) using triangulation, and the dash-dotted lines show the error for a coherent analysis. The thin, medium, and thick lines are for the initial LIGO-Virgo network, the network with Advanced LIGO detectors, and the network with Advanced LIGO detectors and an ``Advanced'' Virgo detector with noise levels comparabale to those of Advanced LIGO detectors, respectively. The lines are terminated at the distance where the SNR from an optimally oriented source is below 5 in all detectors for triangulation, or the network SNR is below 5 for a coherent analysis. The lower dashed line shows the R apparent magnitude (right-hand side axis) expected at peak luminosity from the radioactive decay model. The upper dashed line shows the afterglow luminosity in the shock wave model, 0.1 day after the merger. The top axis shows the expected number of coalescences out to a certain distance.}
\label{fig:nsoptical}
\end{figure}

Figure \ref{fig:nsoptical} presents the position error for a NS-NS binary merger as a function of its distance, the brightness of different R band models, and the rate of coalescence from the Standard model in a spherical volume extending out to that distance, using the extragalactic rate extrapolation method of \citep{Kalogera2001}. 
The position error is a strong function of the position of the source on the sky, and the results of Fig. \ref{fig:nsoptical} are for a source located along the normal to the detectors' plane ($\theta = 0$) and seen face-on.
The right-hand side axis was scaled so that the limiting magnitudes of two optical detectors correspond to their field-of-views.
The first one is ROTSE-III [\citet{ROTSE}, uppermost dotted line in fig. \ref{fig:nsoptical}], a robotic telescope with limiting magnitude $R\sim 18.5$ for 1 minute of integration, a large field-of-view (3.4 deg$^2$), and which can respond in $\sim 10$ s to a trigger.
The second one is the Suprime-Cam camera of the 8.2 m Subaru Telescope (lowest dotted line), which has a large field-of-view (0.25 deg$^2$) for a limiting magnitude of $R \sim 26$ for a 10 minutes exposition\footnote{www.naoj.org/Observing/Instruments/SCam}.

In addition to this figure, Table \ref{tab:results} presents information for a full-sky coverage.
The value quoted in the fourth column is the distance at which the position error for a given source and GW detection system is equal to the EM detector field-of-view (third column), assuming a binary seen face-on along the normal to the detectors' plane.
The fifth column presents the SNR in the EM detector for a source at that distance.
For the ROTSE-III and Subaru telescopes, the radioactive decay model is used for the R band luminosity.
For X-ray observations, I assume the use of the High Resolution Camera on-board the Chandra X-ray Observatory\footnote{hea-www.harvard.edu/HRC}, which has a 0.25 deg$^2$ field-of-view, with a limiting flux sensitivity of $6\times 10^{-14}$ erg cm$^{-2}$ s$^{-1}$ for a 10 minutes integration time.
The EM signal model is then the relativistic blast wave.
Finally, I used the magnetospheric interaction model for radio observations, and assume the use of the Very Large Array telescope; in the P band, it has a 1 mJy RMS sensitivity for 10 minutes of integration, and a 5 deg$^2$ field-of-view (primary beam full width at half power of 2.5 deg)\footnote{www.aoc.nrao.edu/vla/obstatus/vlas/vlas.html}.
I assume below in the rate estimates for radio observations that all NS binaries which coalesce contain at least one NS with a very strong magnetic field. 
Depending on the formation scenario of NS binaries, only a fraction of the binaries may actually contain a strong field NS, and therefore be candidates for radio emission by the magnetospheric interaction model.
\begin{deluxetable}{cccccc}
\tablecolumns{6}
\tablewidth{0pc} 
\tablecaption{Observation prospects for counterparts}
\tablehead{\colhead{source} & \colhead{GW net\tablenotemark{a}} & \colhead{EM instrument} & \colhead{distance} & \colhead{EM SNR} & \colhead{rate} \\
\colhead{} & \colhead{} & \colhead{} & \colhead{(Mpc)} & \colhead{} & \colhead{(yr$^{-1}$)}}
\startdata
NS-NS & I & VLA & 30 & 30 & $2\times 10^{-3}$ \\

BH-NS & I & ROTSE-III & 40 & 100 & $3\times 10^{-3}$ \\

NS-NS & II & ROTSE-III & 60 & $3\times 10^4$ & 0.01 \\

NS-NS & II & VLA & 70 & 4 & 0.02 \\

BH-NS & II & ROTSE-III & 100 & 900 & 0.02 \\

NS-NS & III & Subaru & 100 & $2\times 10^4$ & 0.1 \\

NS-NS & III & Chandra & 100 & $3 \times 10^4$ & 0.1 \\ 

NS-NS & III & ROTSE-III & 400 & 8 & 7 \\

NS-NS & III & VLA & 500 & 0.07 & 10 \\

BH-NS & III & Subaru & 300 & $4\times 10^3$ & 0.2 \\

BH-NS & III & Chandra & 300 & $6\times 10^3$ & 0.2 \\ 

BH-NS & III & ROTSE-III & 900 & 0.4 & 10 \\


\enddata
\tablenotetext{a}{The symbols I, II, and III refer to the initial network, the Advanced LIGO-Virgo network, and the network with Advanced LIGO detectors and an Advanced Virgo detector, respectively. In all cases, a coherent analysis is assumed.}
\label{tab:results}
\end{deluxetable}

For every sky position, binary inclination angle, and GW polarization angle, there is a maximum distance at which the position error box size equals the field-of-view of the EM detector.
Integrating the binary coalescence rate over the whole sky and out to that maximal distance, and averaging over the inclination and polarization angles, gives the expected number of counterparts that will be observable with that EM detector, per unit time.
This rate is quoted in the last column of Table \ref{tab:results}.
It should be noted that the counterpart observation rates do not depend directly on the models of EM emission discussed previously. 
It is only assumed in computing these rates that the EM counterpart will be detectable out to the maximal distance where the error box size equals the field-of-view of the EM detector. 
For the models of EM emission under consideration, the fifth column of Table 1 shows that this condition is nearly always met.

If we happen to observe one NS-NS (NS-BH) merger with the initial network (by luck, or because the rate is larger by a factor of $\sim 100$ than the Standard model), it will have to be within 20 Mpc (40 Mpc) from us in order to allow its observation in the optical wavelengths by a large field-of-view instrument like ROTSE-III, or within 30 Mpc for radio observations, and it will have to be near the normal of the detector's plane.
With Advanced LIGO detectors, the predicted EM counterpart observation rate for NS-NS or BH-NS events is still only one every 25 years.
The required EM detector's field-of-view to achieve a certain detection rate scales like the rate to the 2/3 power\footnote{(detection rate) $\propto$ (distance reach)$^3$, and (detector's field of view) $\sim$ (error box size) $\propto$ (SNR)$^{-2}$ $\propto$ (distance reach)$^2$ $\propto$ (rate)$^{2/3}$}, so EM couterparts observation rates of one per year might be achieved with Advanced LIGO and Initial Virgo detectors if it were possible to effectively detect the EM source in a region of $\sim 40$ deg$^2$.
With a futuristic GW network where Virgo has a noise level comparable to Advanced LIGO instruments, tens of observations of counterparts could be made every year.
Occasional deep observation of binary mergers might also be possible with 8-meter class optical telescopes and the Chandra X-ray observatory.
Given the errors in the estimate of the compact binary merger rate, the predicted rates of counterpart observations could be an order of magnitude larger or smaller.
Consequently, the observation of EM counterparts to compact binary mergers is improbable for the Initial network, possible with Advanced LIGO detectors, and likely with an Advanced detector in Europe.

The optimal exploitation of the world-wide GW network will require the use of coherent analyses, and the development of advanced computational strategies might be necessary for this to be possible.
Fully coherent analyses might be triggerred by less expensive incoherent ones, and therefore be ran only on a small subset of the data.
In most cases, the EM counterparts might be rapidly dimming objects, so that it would be desirable to initiate the observation campaign within a few hours of the GW signal, at most.
It is currently planned that GW data will be analyzed nearly in real-time, and the rapid transfer of data from multiple sites to a central processing facility has already been demonstrated \citep{Marka}.
The development of a world-wide warning system [like SNEWS, \citet{SNEWS}] could also prove essential.
In addition, some sources, like supernovae, might never have a GW waveform known with enough sophistication to perform matched filtering.
It will then be necessary, in order to observe counterparts to these GW events, to develop robust source positioning methods for GW networks [\citet{tinto,CPF}], which do not require a signal model as precise as for matched filtering.

The addition of other sensitive instruments in the GW network, especially in Asia or Oceania, would improve the likelihood of observing EM counterparts, not necessarily by dramatically reducing the size of the best error boxes, but rather by improving the fraction of the sky where the error boxes are small enough to allow EM counterparts observations.
Finally, it might be necessary to get beyond the rough EM models reviewed in this communication in order to optimize the search strategies, and to correctly interpret future observations, or absence thereof.
Large regions of the sky will have to be searched in response to GW triggers, perhaps slightly beyond the capabilities of currently available observatories, especially with respect to the rejection of the astronomical background of variable objects.

\acknowledgments
This work was supported by the National Science Foundation under cooperative agreement PHY-9210038 and the award PHY-0107417.
This document has been assigned LIGO Laboratory document number LIGO-P020024-01-R.

\end{document}